\documentclass[a4paper,showpacs,showkeys,nofootinbib,aps]{revtex4}
\usepackage{amsmath}
\usepackage{epsfig}
\begin{document}
\title{Azimuthal Dependence of the Heavy Quark Initiated Contributions to DIS}
\author{L.N.~Ananikyan}
 \email{lev@web.am}
 \affiliation{Yerevan Physics Institute, Alikhanian Br.2, 375036 Yerevan, Armenia}
\author{N.Ya.~Ivanov}
 \email{nikiv@uniphi.yerphi.am}
\affiliation{Yerevan Physics Institute, Alikhanian Br.2, 375036 Yerevan, Armenia}
\date{\today}
\begin{abstract}
\noindent We analyze the azimuthal dependence of the heavy-quark-initiated contributions to the
lepton-nucleon deep inelastic scattering (DIS). First we derive the relations between the parton
level semi-inclusive structure functions and the helicity $\gamma^{*}Q$ cross sections in the case
of arbitrary values of the heavy quark mass. Then the azimuth-dependent ${\cal O}(\alpha_{s})$
lepton-quark DIS is calculated in the helicity basis. Finally, we investigate numerically the
properties of the $\cos\varphi$ and $\cos2\varphi$ distributions caused by the photon-quark
scattering (QS) contribution. It turns out that, contrary to the basic photon-gluon fusion (GF)
component, the QS mechanism is practically $\cos2\varphi$-independent. This fact implies that
measurements of the azimuthal distributions in charm leptoproduction could directly probe the charm
density in the proton.
\end{abstract}
\pacs{12.38.-t, 13.60.-r, 13.88.+e}%
\keywords{Perturbative QCD, Heavy Flavor Leptoproduction, Intrinsic Charm, Azimuthal Asymmetries}
\maketitle
\section{Introduction}
The notion of the intrinsic charm (IC) content of the proton has been introduced over 25 years ago
in Refs~\cite{BHPS,BPS}. It was shown that, in the light-cone Fock space picture
\cite{brod1,brod2}, it is natural to expect a five-quark state contribution, $\left\vert
uudc\bar{c}\right\rangle$, to the proton wave function. This component can be generated by
$gg\rightarrow c\bar{c}$ fluctuations inside the proton where the gluons are coupled to different
valence quarks. The original concept of the charm density in the proton \cite{BHPS,BPS} has
nonperturbative nature since a five-quark contribution $\left\vert uudc\bar{c}\right\rangle$ scales
as $1/m^{2}$ where $m$ is the $c$-quark mass \cite{polyakov}.

A decade ago another point of view on the charm content of the proton has been proposed in the
framework of the variable flavor number scheme (VFNS) \cite{ACOT,collins}. Within the VFNS, the
mass logarithms of the type $\alpha_{s}\ln\left( Q^{2}/m^{2}\right)$ are resummed through the all
orders into a heavy quark density which evolves with $Q^{2}$ according to the standard DGLAP
evolution equation. Hence this approach introduces the parton distribution functions (PDFs) for the
heavy quarks and changes the number of active flavors by one unit when a heavy quark threshold is
crossed. Note also that the charm density arises within the VFNS perturbatively via the
$g\rightarrow c\bar{c}$ evolution. Some recent developments concerning the VFNS are presented in
Refs.~\cite{Thorne-NNLO,CTEQ6-5HQ,CTEQ6HQ,chi,SACOT}.

Presently, both nonperturbative IC and perturbative charm density are widely used for a
phenomenological description of available data. (A recent review of the theory and experimental
constraints on the charm quark distribution can be found in Refs.~\cite{pumplin,brod-higgs}). In
particular, practically all the recent versions of the CTEQ \cite{CTEQ6} and MRST \cite{MRST2004}
sets of PDFs are based on the VFN schemes and contain a charm density. At the same time, the key
question remains open: How to measure the charm content of the proton? The basic theoretical
problem is that radiative corrections to the leading order (LO) predictions for the heavy quark
production cross sections are large: they increase the Born level results by approximately a factor
of two at energies of the fixed target experiments. On the other hand, perturbative instability
leads to a high sensitivity of the theoretical calculations to standard uncertainties in the input
QCD parameters: $m$, $\mu _{R}$, $\mu _{F}$, $\Lambda _{QCD}$ and PDFs. For this reason, one can
only estimate the order of magnitude of the pQCD predictions for the heavy flavor production cross
sections \cite{Mangano-N-R,Frixione-M-N-R}.

At not very high energies, the main reason for large NLO cross sections of heavy flavor production
in $\gamma g$ \cite{Ellis-Nason,Smith-Neerven}, $\gamma ^{*}g$ \cite{LRSN}, and $gg$
\cite{Nason-D-E-1,Nason-D-E-2,Nason-D-E-3,BKNS} collisions is the so-called threshold (or
soft-gluon) enhancement.  This strong logarithmic enhancement has universal nature in the
perturbation theory since it originates from incomplete cancellation of the soft and collinear
singularities between the loop and the bremsstrahlung contributions. Large leading and
next-to-leading threshold logarithms can be resummed to all orders of perturbative expansion using
the appropriate evolution equations \cite{Contopanagos-L-S,Laenen-O-S,Kidonakis-O-S}. Soft gluon
resummation of the threshold Sudakov logarithms indicates that the higher-order contributions to
the heavy flavor production are also sizeable. (For a review see
Refs.~\cite{Laenen-Moch,kid2,kid1}).

Since production cross sections are not perturbatively stable, it is of special interest to study
those observables that are well-defined in pQCD. A nontrivial example of such an observable was
proposed in Refs.~\cite{we1,we2,we4,we3} where the azimuthal $\cos2\varphi$ asymmetry in heavy
quark photo- and leptoproduction has been analyzed \footnote{The well-known examples are the shapes
of differential cross sections of heavy flavor production which are sufficiently stable under
radiative corrections.}. In particular, the Born level results have been considered \cite{we1,we4}
and the NLO soft-gluon corrections to the basic mechanism, photon-gluon fusion (GF), have been
calculated \cite{we2,we4}. It was shown that, contrary to the production cross sections, the
$\cos2\varphi$ asymmetry in heavy flavor photo- and leptoproduction is quantitatively well defined
in pQCD: the contribution of the dominant GF mechanism to the asymmetry is stable, both
parametrically and perturbatively. This fact provides the motivation for investigation of the
photon-(heavy) quark scattering (QS) contribution to the $\varphi$-dependent lepton-hadron deep
inelastic scattering (DIS).

In the present paper, we calculate the azimuthal dependence of the next-to-leading order (NLO)
${\cal O}(\alpha_{em}\alpha _{s})$ heavy-flavor-initiated contributions to DIS. To our knowledge,
pQCD predictions for the $\varphi $-dependent $\gamma ^{* }Q$ cross sections in the case of
arbitrary values of the heavy quark mass $m$ and $Q^{2}$ are not available in the literature.
Moreover, there is a confusion among the existing results for azimuth-independent $\gamma ^{*}Q$
cross sections.

The NLO corrections to the $\varphi$-independent lepton-quark DIS have been calculated (for the
first time) a long time ago in Ref.~\cite{HM}, and have been re-calculated recently in \cite{KS}.
The authors of Ref.~\cite{KS} conclude that there are errors in the NLO expression for
$\sigma^{(2)}$ given in Ref.~\cite{HM} \footnote{For more details see PhD thesis \cite{KS-thesis},
pp.~158-160.}. We disagree with this conclusion. It will be shown below that a correct
interpretation of the notations for the production cross sections used in \cite{HM} leads to a
complete agreement between the results presented in Refs.~\cite{HM}, \cite{KS} and present paper.

As to the $\varphi $-dependent $\gamma^{*}Q$ cross sections, our main result can be formulated as
follows. Contrary to the basic GF component, the QS mechanism is practically $\cos
2\varphi$-independent. This is due to the fact that the QS contribution to the $\cos 2\varphi $
asymmetry is absent (for the kinematic reason) at LO and is negligibly small (of the order of
$1\%$) at NLO. This fact indicates that the azimuthal distributions in charm leptoproduction could
be a good probe of the charm density in the proton. In detail, the possibility of measuring the
charm content of the proton using the $\cos 2\varphi$ asymmetry will be investigated in a
forthcoming publication \cite{we5}.

Concerning the experimental aspects, azimuthal asymmetries in charm leptoproduction can, in
principle, be measured in the COMPASS experiment at CERN, as well as in future studies at the
proposed eRHIC \cite{eRHIC,EIC} and LHeC \cite{LHeC} colliders at BNL and CERN, correspondingly.

The outline of this paper is as follows. In Section~\ref{I}, we derive the relations between the
parton level semi-inclusive structure functions and the helicity $\gamma^{*}Q$ cross sections in
the case of arbitrary values of the heavy quark mass. As explained in Ref.~\cite{AOT}, in the
presence of non-zero masses, it is the helicity basis that provides the simplest connections
between the hadron- and parton-level production cross sections. In Section~\ref{II}, we present the
NLO ${\cal O}(\alpha_{em}\alpha _{s})$ predictions for the $\varphi$-dependent lepton-quark DIS in
the helicity basis. Our calculations are compared with available results in Section~\ref{III}. In
Section~\ref{IV}, a numerical investigation of the $\cos\varphi$ and $\cos2\varphi$ distributions
caused by the QS contribution is given. In particular, we provide a simple parton level proof of
the fact that the QS mechanism is practically $\cos 2\varphi$-independent. Our conclusions are
presented in Section~\ref{V}.

\section{\label{I}Azimuth-Dependent Structure Functions in the Helicity Basis}

In this Section, the helicity formalism for the semi-inclusive $\gamma ^{*}Q$ cross sections in the
case of arbitrary values of the heavy quark mass is presented. This is a purely kinematical
analysis, which will set the notation to be used later on. In fact, we extend the helicity approach
proposed in Ref.~\cite{AOT} to the case of $\varphi$-dependent leptoproduction using the method
formulated in Ref.~\cite{dombey}.

\begin{figure}
\begin{center}
\mbox{\epsfig{file=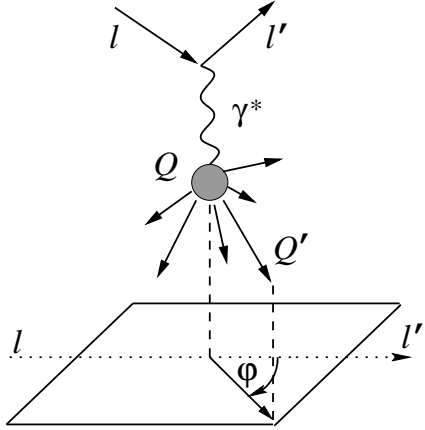,width=200pt}} \caption{\label{Fg.1}\small Definition of the
azimuthal angle $\varphi$ in the target rest frame.}
\end{center}
\end{figure}

We consider the semi-inclusive deep inelastic lepton-quark scattering. The momentum assignment will
be denoted as
\begin{equation}\label{1}
l(\ell )+Q(k)\rightarrow l^{\prime}(\ell -q)+Q^{\prime}(p)+X(p_{X}).
\end{equation}
The following definition of partonic kinematic variables is used:
\begin{equation}\label{2}
y=\frac{q\cdot k}{\ell \cdot k},\qquad \qquad z=\frac{Q^{2}}{2q\cdot k} ,\qquad \qquad \lambda
=\frac{m^{2}}{Q^{2}},\qquad \qquad Q^{2}=-q^{2}.
\end{equation}
The differential cross section of the reaction (\ref{1}), d$^{3}\hat{\sigma} _{lQ}$, is defined in
terms of the quark tensor $\widetilde{W} _{Q}^{\mu \nu }$:
\begin{equation}\label{3}
\ell _{0}^{\prime }\frac{\text{d}^{3}\hat{\sigma}_{lQ}}{\text{d}^{3}\ell ^{\prime
}}=2\frac{\text{d}^{3}\hat{\sigma}_{lQ}}{\text{d}y\text{d}Q^{2} \text{d}\varphi
}=\frac{2z}{y}\frac{\text{d}^{3}\hat{\sigma}_{lQ}}{\text{d}z \text{d}Q^{2}\text{d}\varphi
}=\frac{\alpha _{em}^{2}}{(\ell \cdot k)Q^{4}}L_{\mu \nu }\widetilde{W}_{Q}^{\mu \nu
}\frac{\text{d}^{3}p}{(2\pi )^{3}2p_{0}},
\end{equation}
where $\ell^{\prime}_{\mu}=(\ell -q)_{\mu}$ is the 4-momentum of the final lepton. In the target
rest frame, the azimuth $\varphi $ is the angle between the lepton scattering plane and the heavy
quark production plane, defined by the exchanged photon and the detected quark $Q^{\prime}$ (see
Fig.~\ref{Fg.1}). The covariant definition of $\varphi $ is
\begin{eqnarray}
\cos \varphi &=&\frac{r\cdot n}{\sqrt{-r^{2}}\sqrt{-n^{2}}},\qquad \qquad \qquad \qquad \sin
\varphi =\frac{Q^{2}\sqrt{1+4\lambda z^{2}}}{2z\sqrt{-r^{2}}\sqrt{-n^{2}
}}~n\cdot \ell , \nonumber\\
r^{\mu } &=&\varepsilon ^{\mu \nu \alpha \beta }k_{\nu }q_{\alpha }\ell _{\beta },\qquad \qquad
\qquad \qquad \quad n^{\mu }=\varepsilon ^{\mu \nu \alpha \beta }q_{\nu }k_{\alpha }p_{\beta
}.\label{4}
\end{eqnarray}
The explicit expression for the lepton tensor $L_{\mu \nu }$ is:
\begin{equation}\label{5}
L_{\mu \nu }=\mathop{\overline{\sum}}\limits_{\text{spin}}\left\langle \ell \left\vert j_{\nu
}^{\dagger }\right\vert \ell ^{\prime }\right\rangle \left\langle \ell ^{\prime }\left\vert j_{\mu
}\right\vert \ell \right\rangle =2\ell _{\mu }\ell _{\nu }^{\prime }+2\ell _{\nu }\ell _{\mu
}^{\prime }-Q^{2}g_{\mu \nu },
\end{equation}
where $\mathop{\overline{\sum}}\limits_{\text{spin}}$denotes a sum over all final helicity states
and an averaging over all initial spin variables. The semi-inclusive quark tensor
$\widetilde{W}_{Q}^{\mu \nu }$ is defined as follows:
\begin{equation}\label{6}
\widetilde{W}_{Q}^{\mu \nu }(q,k,p)=\frac{1}{4\pi }\mathop{\overline{\sum}}\limits
_{X(p_{X}),\text{spin}}\left\langle k\left\vert J^{\mu }\right\vert p,p_{X}\right\rangle \left(
2\pi \right) ^{4}\delta ^{(4)}\left( q+k-p-p_{X}\right) \left\langle p_{X},p\left\vert J^{\dagger
\nu }\right\vert k\right\rangle,
\end{equation}
where sums and integrals over all the unobserved final states $X$ of momentum $p_{X}^{\mu }$ are
implied.

To construct the parton tensor describing the semi-inclusive $\gamma ^{*}Q$ DIS, it is convenient
to introduce two 4-vectors:
\begin{equation}\label{7}
v^{\mu }=\frac{\varepsilon _{\tau \nu \alpha \beta }\varepsilon ^{\tau \gamma \theta \mu }q^{\nu
}k^{\alpha }p^{\beta }k_{\gamma }q_{\theta }}{ (q\cdot k)\sqrt{-n^{2}}\sqrt{1+4\lambda
z^{2}}},\qquad \qquad \qquad \qquad \qquad \qquad w^{\mu }=k^{\mu }+\frac{q\cdot k}{Q^{2}}q^{\mu },
\end{equation}
that obey the following conditions: $v\cdot k=v\cdot q=0,$ $v^{2}=-1$ and $ w\cdot q=0,$ $w\cdot
k=w^{2}=m^{2}\left( 1+4\lambda z^{2}\right) /4\lambda z^{2}$. In terms of $v^{\mu }$ and $w^{\nu
}$, $\widetilde{W}_{Q}^{\mu \nu }$ has the following structure:
\begin{equation}\label{8}
\widetilde{W}_{Q}^{\mu \nu }(q,k,p)=-\left( g^{\mu \nu }-\frac{q^{\mu }q\nu }{q^{2}}\right)
\widetilde{W}_{1}+\frac{w^{\mu }w^{\nu }}{m^{2}}\widetilde{W} _{2}+\left( w^{\mu }v^{\nu }+w^{\nu
}v^{\mu }\right) \frac{\widetilde{W}_{I} }{m}+v^{\mu }v^{\nu }\widetilde{W}_{A},
\end{equation}
that obeys all the necessary conservation laws. In particular, $\widetilde{W} _{Q}^{\mu \nu }q_{\mu
}=0$. The scalar coefficients $\widetilde{W}_{i}$ $(i=1,2,I,A)$ are the semi-inclusive parton-level
structure functions for the process (\ref{1}), $\widetilde{W}_{i}\equiv \widetilde{W}_{i}\left(
z,\lambda ,p_{X}^{2},q\cdot p\right)$.

For parton-model considerations, is it convenient to use the so-called colliner frames where the
3-momenta of the virtual photon and initial quark are antiparallel to each other, $\vec{q}\parallel
(-\vec{k})$. Evidently, an arbitrary colliner frame can be obtained from the initial quark rest
system with the help of a Lorentz boost along $\vec{q}$. Pointing the $z$-axis along $\vec{q}$, we
will have in a colliner frame:
\begin{eqnarray}
v^{\mu } &=&(0,\vec{v}_{\perp },0),\qquad \qquad \qquad \qquad\qquad \qquad  \vec{v}_{\perp
}=\frac{\vec{p}_{\perp }}{\left\vert
\vec{p}_{\perp }\right\vert }, \label{9}\\
w^{\mu } &=&\frac{Q\sqrt{1+4\lambda z^{2}}}{2\sqrt{z}}e_{0}^{\mu },\qquad \qquad \quad \qquad
\qquad\;\, e_{0}^{\mu }=\frac{1}{Q}(\left\vert \vec{q} \right\vert ,\vec{0}_{\perp
},q_{0}),\label{10}
\end{eqnarray}
where $Q=\sqrt{-q^{2}}$ and $e_{0}^{\mu }$ describes the longitudinal polarization of the virtual
photon, $\gamma ^{* }$. It is also useful to define the scalar, $e_{q}^{\mu }$, and transverse,
$e_{\pm }^{\mu }$, polarization vectors:
\begin{equation}\label{11}
e_{q}^{\mu }=\frac{q^{\mu }}{Q},\qquad \qquad \qquad \qquad \qquad \qquad \qquad e_{\pm }^{\mu
}=\frac{1}{\sqrt{2}}\left(0,\mp 1,-i,0\right).
\end{equation}
Note the completeness relation
\begin{equation}\label{12}
e_{+}^{\mu }e_{+}^{\nu * }+e_{-}^{\mu }e_{-}^{\nu * }-e_{0}^{\mu }e_{0}^{\nu * }+e_{q}^{\mu
}e_{q}^{\nu * }=-g^{\mu \nu },
\end{equation}
and the normalization for the physical states:
\begin{equation}\label{13}
e_{r}\cdot e_{s}^{* }=(-1)^{s}\delta_{rs},\qquad \qquad \qquad (r,s=0,\pm 1).
\end{equation}
One can see from Eqs.~(\ref{9},\ref{10}) that it is merely the scalar coefficient functions
$\widetilde{W}_{i}$ $(i=1,2,I,A)$\ depend on the final quark momentum $p$ in a collinear frame. For
this reason, we can integrate the semi-inclusive quark tensor $\widetilde{W}_{Q}^{\mu \nu }(q,k,p)$
over $ \vec{p}$ and obtain the inclisive quantity $W_{Q}^{\mu \nu }(q,k)$:
\begin{eqnarray}
W_{Q}^{\mu \nu }(q,k) &=&\int \frac{\text{d}^{3}p}{(2\pi )^{3}2p_{0}}
\widetilde{W}_{Q}^{\mu \nu }(q,k,p) \label{14}\\
&=&-\left( g^{\mu \nu }-\frac{q^{\mu }q\nu }{q^{2}}\right) \left(\hat{W}
_{1}-\frac{1}{2}\hat{W}_{A}\right) +\frac{w^{\mu }w^{\nu }}{m^{2}}\left(\hat{W}_{2}-\frac{2\lambda
z^{2}}{1+4\lambda z^{2}}\hat{W}_{A}\right) +\left( w^{\mu }v^{\nu }+w^{\nu }v^{\mu }\right)
\frac{\hat{W}_{I}}{m} +v^{\mu }v^{\nu }\hat{W}_{A}.\nonumber
\end{eqnarray}
The inclusive coefficient functions $W_{i}(z,\lambda )$ $\ (i=1,2,I,A)$ are related to the
semi-inclusive ones as follows:
\begin{eqnarray}
\hat{W}_{1}(z,\lambda ) &=&\int \frac{\text{d}^{3}p}{(2\pi )^{3}2p_{0}} \left(
\widetilde{W}_{1}+\frac{1}{2}\widetilde{W}_{A}\right) \left( z,\lambda ,p_{X}^{2},q\cdot p\right)
,\qquad \qquad \; \hat{W} _{I}(z,\lambda )=\int \frac{\text{d}^{3}p}{(2\pi
)^{3}2p_{0}}\widetilde{W}_{I}\left( z,\lambda ,p_{X}^{2},q\cdot p\right) , \label{15}\\
\hat{W}_{2}(z,\lambda ) &=&\int \frac{\text{d}^{3}p}{(2\pi )^{3}2p_{0}} \left(
\widetilde{W}_{2}+\frac{2\lambda z^{2}}{1+4\lambda z^{2}}\widetilde{W} _{A}\right) \left( z,\lambda
,p_{X}^{2},q\cdot p\right),\quad \! \hat{W} _{A}(z,\lambda )=\int \frac{\text{d}^{3}p}{(2\pi
)^{3}2p_{0}}\widetilde{W} _{A}\left( z,\lambda ,p_{X}^{2},q\cdot p\right).\nonumber
\end{eqnarray}
Integrating $W_{Q}^{\mu \nu }(q,k)$ over the lepton azimuth $\varphi $ defined by Eqs.~(\ref{4}),
one can reproduce the well-known expression for totally inclusive DIS:
\begin{equation}\label{16}
\frac{1}{2\pi }\int\limits_{0}^{2\pi }\text{d}\varphi W_{Q}^{\mu \nu }(q,k)=-\left( g^{\mu \nu
}-\frac{q^{\mu }q\nu }{q^{2}}\right) \hat{W} _{1}(z,Q^{2})+\frac{w^{\mu }w^{\nu
}}{m^{2}}\hat{W}_{2}(z,Q^{2}).
\end{equation}
Note that the above relation can easily be obtained from Eq.~(\ref{14}) taking into account that
$\int_{0}^{2\pi }$d$\varphi\,v^{\mu }v^{\nu }=\pi \left( e_{+}^{\mu }e_{+}^{\nu * }+e_{-}^{\mu
}e_{-}^{\nu * }\right) =\pi \left( -g^{\mu \nu }+e_{0}^{\mu }e_{0}^{\nu * }-e_{q}^{\mu }e_{q}^{\nu
* }\right).$

Now the cross section for the inclusive azimuth-dependent lepton-quark DIS can be written as
\begin{equation}\label{17}
\frac{\text{d}^{3}\hat{\sigma}_{lQ}}{\text{d}z\text{d}Q^{2}\text{d}\varphi }=
\frac{y}{z}\frac{\text{d}^{3}\hat{\sigma}_{lQ}}{\text{d}y\text{d}Q^{2}\text{d }\varphi
}=\frac{y}{Q^{2}}\frac{\text{d}^{3}\hat{\sigma}_{lQ}}{\text{d}z\text{d}y\text{d}\varphi
}=\frac{\alpha _{em}^{2}y^{2}}{Q^{6}}L_{\mu \nu }W_{Q}^{\mu \nu }.
\end{equation}
To derive the relations between the invariant and helicity structure functions, we use the
completeness (\ref{12}) which implies that
\begin{equation}\label{18}
L_{\mu \nu }W_{Q}^{\mu \nu }=\sum\limits_{r,s}\rho _{rs}\hat{F}_{rs},
\end{equation}
where the quark and lepton helicity structure functions ($\hat{F}_{rs}$ and $\rho _{rs}$,
respectively) are defined as
\begin{equation}\label{19}
\hat{F}_{rs}=e_{r}^{\mu }W_{Q,\mu \nu }e_{s}^{\nu *},\qquad \qquad \quad \ \ \qquad \qquad \ \rho
_{rs}=(-1)^{r+s}e_{r}^{\mu * }L_{\mu \nu }e_{s}^{\nu }.
\end{equation}
Choosing the $x$-axis along $\vec{v}_{\perp }$ defined by Eq.~(\ref{9}), we obtain for the quark
helicity structure functions $\hat{F}_{rs}(z,\lambda )$:
\begin{eqnarray}
\hat{F}_{{+}{+}} &=&\hat{F}_{{-}{-}}=\hat{W}_{1},\qquad \qquad \quad \ \ \ \qquad \qquad \quad
\quad \;\; \ \ \ \qquad \qquad \quad \ \ \ \ \hat{F}_{\mathop{0}\mathop{0}}=-\hat{W}_{1}+
\frac{1+4\lambda z^{2}}{4\lambda z^{2}}\hat{W}_{2}, \nonumber\\
\hat{F}_{{+}\mathop{0}}&=&\hat{F}_{\mathop{0}{+}}=-\hat{F}_{{-}\mathop{0}}=-\hat{F}_{\mathop{0}{-}}=
\frac{1}{2}\sqrt{\frac{1+4\lambda z^{2}}{2\lambda z^{2}}}\hat{W}_{I},\qquad \qquad \qquad \,
\hat{F} _{{+-}}=\hat{F}_{{-+}}=-\frac{1}{2}\hat{W}_{A}.\label{20}
\end{eqnarray}
The lepton tensor $\rho _{rs}\left( \hat{\varepsilon},\varphi \right)$ has the following form in
the helicity basis:
\begin{eqnarray}
\rho_{{++}} &=&\rho _{{--}}=\frac{Q^{2}}{1-\hat{\varepsilon}},\qquad \qquad \quad \ \ \ \qquad
\qquad \quad \ \ \ \qquad \qquad \quad \ \ \ \ \ \ \, \rho _{\mathop{0}\mathop{0}}=
\frac{2Q^{2}\hat{\varepsilon}}{1-\hat{\varepsilon}}, \label{21}\\
\rho_{{+}\mathop{0}}
&=&\rho_{\mathop{0}{+}}^{*}=-\rho_{{-}\mathop{0}}^{*}=-\rho_{\mathop{0}{-}}=-\frac{Q^{2}
\sqrt{\hat{\varepsilon}(1+\hat{\varepsilon})}}{1-\hat{\varepsilon}} \,\text{e}^{-i\varphi },\qquad
\qquad \ \ \ \, \rho_{{+-}}=\rho_{{-+}}^{* }=-\frac{Q^{2}\hat{
\varepsilon}}{1-\hat{\varepsilon}}\,\text{e}^{-2i\varphi}.\nonumber
\end{eqnarray}
The quantity $\hat{\varepsilon}$ measures the degree of the longitudinal polarization of the
virtual photon in the Breit frame \cite{dombey}. The covariant definition is:
\begin{equation}\label{22}
\hat{\varepsilon}=\frac{2(1-y-\lambda z^{2}y^{2})}{1+(1-y)^{2}+2\lambda z^{2}y^{2}}.
\end{equation}
In terms of the helicity structure functions, the azimuth-dependent inclusive lepton-quark cross
section has the form:
\begin{equation}\label{23}
\frac{\text{d}^{3}\hat{\sigma}_{lQ}}{\text{d}z\text{d}Q^{2}\text{d}\varphi }= \frac{\alpha
_{em}^{2}}{Q^{4}}\frac{2y^{2}}{1-\hat{\varepsilon}}\left[ \hat{F}_{T}(z,\lambda
)+\hat{\varepsilon}\hat{F}_{L}(z,\lambda )+\hat{\varepsilon}\hat{F}_{A}(z,\lambda )\cos 2\varphi
+2\sqrt{\hat{\varepsilon}(1+\hat{ \varepsilon})}\,\hat{F}_{I}(z,\lambda )\cos \varphi \right] ,
\end{equation}
where
\begin{equation}\label{24}
\hat{F}_{T}=\hat{F}_{{++}},\qquad \qquad \hat{F}_{L}=\hat{F}_{\mathop{0}\mathop{0}},\qquad \qquad
\hat{F}_{A}=-\hat{F}_{{+-}},\qquad \qquad \hat{F}_{I}=\hat{F}_{{-}\mathop{0}}.
\end{equation}
Likewise, using Eqs.~(\ref{18}-\ref{22}), one can easily express the semi-inclusive $lQ$ cross
section defined by Eq.~(\ref{3}) in terms of the corresponding helicity structure functions
$\widetilde{F}_{rs}\left( z,\lambda ,k\cdot p,q\cdot p\right)=e_{r}^{\mu }(q,k)\widetilde{W}_{Q,\mu
\nu }(q,k,p)e_{s}^{\nu *}(q,k)$.

Sometimes, instead of the structure functions $\hat{F}_{rs}$, the helicity $\gamma ^{* }Q$ cross
sections are used:
\begin{equation}\label{25}
\hat{\sigma}_{i}(z,\lambda )=\frac{8\pi ^{2}\alpha _{em}\,z}{Q^{2}\sqrt{ 1+4\lambda
z^{2}}}\,\hat{F}_{i}(z,\lambda),\qquad (i=T,L,A,I),
\end{equation}
where $\hat{\sigma}_{T}=\hat{\sigma}_{{++}},\
\hat{\sigma}_{L}=\hat{\sigma}_{\mathop{0}\mathop{0}},\ \hat{\sigma}_{A}=-\hat{\sigma}_{{+-}}$ and
$\hat{\sigma}_{I}=\hat{\sigma}_{{-}\mathop{0}}$. Since $y\ll 1$ in most of the experimentally
reachable kinematic range, it is the the quantities $\hat{\sigma}_{2}$ and $\hat{F}_{2}$ that can
effectively be measured in $\varphi$-independent DIS:
\begin{equation}\label{26}
\hat{\sigma}_{2}(z,\lambda)=\hat{\sigma}_{T}(z,\lambda)+\hat{\sigma}_{L}(z,\lambda),\qquad \qquad
\qquad \qquad \hat{F}_{2}(z,\lambda)= \frac{2z}{1+4\lambda
z^{2}}\left[\hat{F}_{T}(z,\lambda)+\hat{F}_{L}(z,\lambda)\right].
\end{equation}
In terms of the quantities $\hat{\sigma}_{i}$, the cross section of the reaction (\ref{1}) can be
written as
\begin{equation}\label{27}
\frac{\text{d}^{3}\hat{\sigma}_{lQ}}{\text{d}z\text{d}Q^{2}\text{d}\varphi }= \frac{\alpha
_{em}}{(2\pi )^{2}}\frac{y^{2}}{zQ^{2}}\frac{\sqrt{1+4\lambda z^{2}}}{1-\hat{\varepsilon}}\left[
\hat{\sigma}_{2}(z,\lambda )-(1-\hat{ \varepsilon})\hat{\sigma}_{L}(z,\lambda
)+\hat{\varepsilon}\hat{\sigma}_{A}(z,\lambda )\cos 2\varphi
+2\sqrt{\hat{\varepsilon}(1+\hat{\varepsilon})}\,\hat{\sigma}_{I}(z,\lambda )\cos \varphi \right].
\end{equation}
In Eqs.~(\ref{26},\ref{27}), $\hat{\sigma}_{T}\,(\hat{\sigma}_{L})$ is the usual $ \gamma ^{* }N$
cross section describing heavy quark production by a transverse (longitudinal) virtual photon. The
third cross section, $\hat{ \sigma}_{A}$, comes about from interference between transverse states
and is responsible for the $\cos 2\varphi $ asymmetry which occurs in real photoproduction using
linearly polarized photons \cite{we1,we2,we3}. The fourth cross section, $\hat{\sigma}_{I}$,
originates from interference between longitudinal and transverse components \cite{dombey}.

\section{\label{II}Photon-Quark Scattering Cross Sections at NLO}
\begin{figure}
\begin{center}
\mbox{\epsfig{file=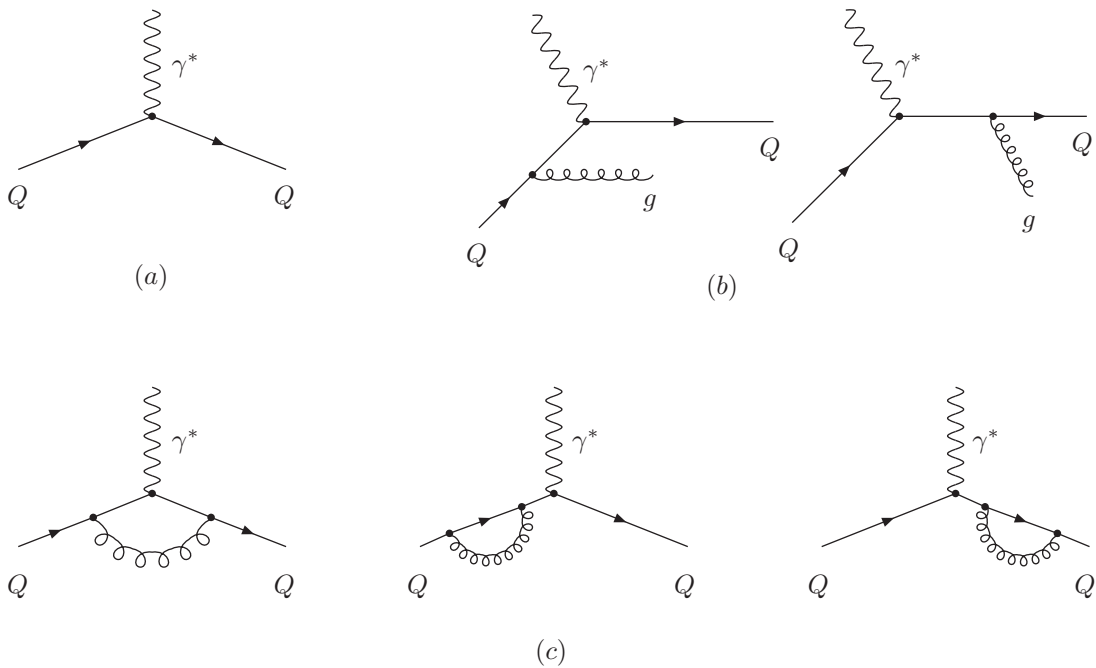,width=445pt}}
\end{center}
\caption{\label{Fg.2}\small The LO (a) and NLO (b and c) photon-quark scattering diagrams.}
\end{figure}

At leading order, ${\cal O}(\alpha _{em})$, the only quark scattering subprocess is
\begin{equation}
\gamma ^{*}(q)+Q(k_{Q})\rightarrow Q(p_{Q}).  \label{21.2}
\end{equation}
The $\gamma ^{*}Q$ cross sections, $\hat{\sigma}_{i}^{(0)}$ ($i=2,L,A,I$), corresponding to the
Born diagram (see Fig.~\ref{Fg.2}a) are:
\begin{eqnarray}
\hat{\sigma}_{2}^{(0)}(z,\lambda)&=&\hat{\sigma}_{B}(z)\sqrt{1+4\lambda z^{2}}\,\delta(1-z), \nonumber\\
\hat{\sigma}_{L}^{(0)}(z,\lambda)&=&\hat{\sigma}_{B}(z)\frac{4\lambda z^{2}}{\sqrt{1+4\lambda
z^{2}}}\,\delta(1-z), \label{22.2}\\
\hat{\sigma}_{A}^{(0)}(z,\lambda)&=&\hat{\sigma}_{I}^{(0)}(z,\lambda)=0,\nonumber
\end{eqnarray}
with
\begin{equation}\label{23.2}
\hat{\sigma}_{B}(z)=\frac{(2\pi)^2e_{Q}^{2}\alpha _{em}}{Q^{2}}\,z,
\end{equation}
where $e_{Q}$ is the quark charge in units of electromagnetic coupling constant.

To take into account the NLO ${\cal O}(\alpha_{em}\alpha_{s})$ contributions, one needs to
calculate the virtual corrections to the Born process  (given in Fig.~\ref{Fg.2}c) as well as the
real gluon emission (see Fig.~\ref{Fg.2}b):
\begin{equation}
\gamma ^{*}(q)+Q(k_{Q})\rightarrow Q(p_{Q})+g(p_{g}).  \label{24.2}
\end{equation}

The NLO $\varphi$-dependent cross sections, $\hat{\sigma}_{A}^{(1)}$ and $\hat{\sigma}_{I}^{(1)}$,
are described by the real gluon emission only. Corresponding contributions are free of any type of
singularities and the quantities $\hat{\sigma}_{A}^{(1)}$ and $\hat{\sigma}_{I}^{(1)}$ can be
calculated directly in four dimensions.

In the $\varphi$-independent case, $\hat{\sigma}_{2}^{(1)}$ and $\hat{\sigma}_{L}^{(1)}$, we also
work in four dimensions. The virtual contribution (Fig.~\ref{Fg.2}c) contains ultraviolet (UV)
singularity that is removed using the on-mass-shell regularization scheme. In particular, we
calculate the absorptive part of the Feynman diagram which has no UV divergences.  The real part is
then obtained by using the appropriate dispersion relations. As to the infrared (IR) singularity,
it is regularized with the help of an infinitesimal gluon mass. This IR divergence is cancelled
when we add the bremsstrahlung contribution (Fig.~\ref{Fg.2}b).

The final (real+virtual) results for $\gamma ^{*}Q$ cross sections can be cast into the following
form:
\begin{eqnarray}
\hat{\sigma}_{2}^{(1)}(z,\lambda)=\frac{\alpha_{s}}{2\pi}C_{F}\hat{\sigma}_{B}(1)
\sqrt{1+4\lambda}\,\delta(1-z)\Bigl\{-2+4\ln\lambda-\sqrt{1+4\lambda }\,\ln r+
\frac{1+2\lambda}{\sqrt{1+4\lambda}}\Bigl[2\text{Li}_{2}(r^{2})+4\text{Li}_{2}(-r)&& \nonumber\\
+3\ln^{2}(r)-4\ln r+4\ln r \ln(1+4\lambda)-2\ln r\ln\lambda\Bigr]\Bigr\}&& \nonumber \\
+\frac{\alpha_{s}}{4\pi}C_{F}\hat{\sigma}_{B}(z)\frac{1}{(1+4\lambda z^{2})^{3/2}}\biggl\{
\frac{1}{\left[1-(1-\lambda)z\right]^{2}}\Bigl[1-3z-4z^{2}+6z^{3}+8z^{4}-8z^{5} \qquad \qquad \qquad \;&& \nonumber \\
+6\lambda z\left(3-18z+13z^{2}+10z^{3}-8z^{4}\right) \qquad \qquad&&  \nonumber \\
+4\lambda^{2}z^{2}\left(8-77z+65z^{2}-2z^{3}\right)\biggr. \qquad \qquad \qquad \; \,&&  \label{25.2} \\
+16\lambda^{3}z^{3}\left(1-21z+12z^{2}\right)-128\lambda^{4}z^{5}\Bigr] \qquad \quad \; \,&&   \nonumber \\
+\frac{2\ln D(z,\lambda)}{\sqrt{1+4\lambda z^{2}}}\Bigl[-\left(1+z+2z^{2}+2z^{3}\right)+2\lambda
z\left(2-11z-11z^{2}\right)+8\lambda^{2}z^{2}\left(1-9z\right)\Bigr]&& \nonumber \\
-\frac{8(1+4\lambda)^{2}z^{4}}{\left(1-z\right)_{+}}-
\frac{4(1+2\lambda)(1+4\lambda)^{2}z^{4}}{\sqrt{1+4\lambda z^{2}}}\frac{\ln
D(z,\lambda)}{\left(1-z\right)_{+}}\biggr\},&& \nonumber
\end{eqnarray}
\begin{eqnarray}
\hat{\sigma}_{L}^{(1)}(z,\lambda)=\frac{\alpha_{s}}{\pi}C_{F}\hat{\sigma}_{B}(1)
\frac{2\lambda}{\sqrt{1+4\lambda}}\delta(1-z)\Bigl\{-2+4\ln\lambda-\frac{4\lambda}{\sqrt{1+4\lambda
}}\,\ln r+\frac{1+2\lambda}{\sqrt{1+4\lambda}}\Bigl[2\text{Li}_{2}(r^{2})+4\text{Li}_{2}(-r)&& \nonumber\\
+3\ln^{2}(r)-4\ln r+4\ln r \ln(1+4\lambda)-2\ln r\ln\lambda\Bigr]\Bigr\}&& \nonumber \\
+\frac{\alpha_{s}}{\pi}C_{F}\hat{\sigma}_{B}(z)\frac{1}{(1+4\lambda z^{2})^{3/2}}\biggl\{
\frac{z}{\left[1-(1-\lambda)z\right]^{2}}\Bigl[(1-z)^{2}-
\lambda z\left(13-19z-2z^{2}+8z^{3}\right)\Bigr. \qquad \qquad&&   \nonumber \\
-2\lambda^{2}z^{2}\left(31-39z+8z^{2}\right)\Bigr. \qquad \qquad \qquad \qquad  \quad \; \,&& \label{26.2} \\
-8\lambda^{3}z^{3}\left(10-7z\right)-32\lambda^{4}z^{4}\Bigr]\Bigr.  \qquad \qquad \qquad  \qquad&&   \nonumber \\
-\frac{2\lambda z^{2}\ln D(z,\lambda)}{\sqrt{1+4\lambda z^{2}}}\left[3+3z+16\lambda z\right]\Bigr.
\qquad \qquad \qquad \qquad \; \,&& \nonumber \\
-\frac{8\lambda(1+4\lambda)z^{4}}{\left(1-z\right)_{+}}-
\frac{4\lambda(1+2\lambda)(1+4\lambda)z^{4}}{\sqrt{1+4\lambda z^{2}}}\frac{\ln
D(z,\lambda)}{\left(1-z\right)_{+}}\biggr\},&& \nonumber
\end{eqnarray}
\begin{equation}\label{27.2}
\hat{\sigma}_{A}^{(1)}(z,\lambda)=\frac{\alpha_{s}}{2\pi}C_{F}\hat{\sigma}_{B}(z)\frac{z(1-z)}{(1+4\lambda
z^{2})^{3/2}}\biggl\{ \frac{1}{\left[1-(1-\lambda)z\right]}\left[1+2\lambda(4-3z)+8\lambda^2
z\right]+\frac{2\lambda \ln D(z,\lambda)}{\sqrt{1+4\lambda z^{2}}}\left[2+z+4\lambda
z\right]\biggr\},
\end{equation}
\begin{eqnarray}
\hat{\sigma}_{I}^{(1)}(z,\lambda)=\frac{\alpha_{s}}{8\sqrt{2}}C_{F}\hat{\sigma}_{B}(z)
\frac{1}{(1+4\lambda z^{2})^{2}}\frac{\sqrt{z}}{\left[1-(1-\lambda)z\right]^{3/2}}\biggl\{
-(1-z)(1+2z)-4\lambda z\left(10-10z-z^{2}+2z^{3}\right)  \qquad \quad \; \;&& \nonumber \\
-8\lambda^{2}z^{2}\left(25-29z+8z^{2}\right)-96\lambda^{3}z^{3}\left(3-2z\right)-
128\lambda^{4}z^{4}\biggr.\;&& \label{28.2} \\
+8\sqrt{\lambda z\left[1-(1-\lambda)z\right]}\left[1-z^{2}+\lambda
z(13-11z)+4\lambda^{2}z^{2}(7-4z)+16\lambda^{3}z^{3}\right]\biggr\}.&& \nonumber
\end{eqnarray}
In Eqs.~(\ref{25.2}-\ref{28.2}), $C_{F}=(N_{c}^{2}-1)/(2N_{c})$, where $N_{c}$ is number of colors,
while
\begin{equation}\label{29.2}
D(z,\lambda)=\frac{1+2\lambda z -\sqrt{1+4\lambda z^{2}}}{1+2\lambda z +\sqrt{1+4\lambda
z^{2}}},\qquad \qquad \qquad \qquad \qquad
r=\sqrt{D(z=1,\lambda)}=\frac{\sqrt{1+4\lambda}-1}{\sqrt{1+4\lambda}+1}.
\end{equation}
The so-called "plus" distributions are defined by
\begin{equation}\label{30.2}
\left[g(z)\right]_{+}=g(z)-\delta(1-z)\int\limits_{0}^{1}\text{d}\zeta\,g(\zeta).
\end{equation}
For any sufficiently regular test function $h(z)$, Eq.~(\ref{30.2}) gives
\begin{equation}\label{31.2}
\int\limits_{a}^{1}\text{d}z\,h(z)\left[\frac{\ln^{k}(1-z)}{1-z}\right]_{+}=
\int\limits_{a}^{1}\text{d}z\frac{\ln^{k}(1-z)}{1-z}\left[h(z)-h(1)\right]+
h(1)\frac{\ln^{k+1}(1-a)}{k+1}.
\end{equation}

\section{\label{III}Comparison with Available Results}
For the first time, the NLO ${\cal O}(\alpha_{em}\alpha_{s})$ corrections to the
$\varphi$-independent IC contribution have been calculated a long time ago by Hoffmann and Moore
(HM) \cite{HM}. However, authors of Ref.~\cite{HM} don't give explicitly their definition of the
partonic cross sections that leads to a confusion in interpretation of the original HM results. To
clarify the situation, we need first to derive the relation between the lepton-quark DIS cross
section, $\text{d}\hat{\sigma}_{lQ}$, and the partonic cross sections, $\sigma^{(2)}$ and
$\sigma^{(L)}$, used in \cite{HM}. Using Eqs.~(C.1) and (C.5) in Ref.~\cite{HM}, one can express
the HM tensor $\sigma_{R}^{\mu\nu}$ in terms of "our" cross sections $\hat{\sigma}_{2}$ and
$\hat{\sigma}_{L}$ defined by Eq.~(\ref{27}) in the present paper. Comparing the obtained results
with the corresponding definition of $\sigma_{R}^{\mu\nu}$ via the HM cross sections $\sigma^{(2)}$
and $\sigma^{(L)}$ (given by Eqs.~(C.16) and (C.17) in Ref.~\cite{HM}), we find that
\begin{eqnarray}
\hat{\sigma}_{2}(z,\lambda)&\equiv &\hat{\sigma}_{B}(z)\sqrt{1+4\lambda
z^{2}}\,\sigma^{(2)}(z,\lambda),  \label{35.3} \\
\hat{\sigma}_{L}(z,\lambda)&\equiv &\frac{2\hat{\sigma}_{B}(z)}{\sqrt{1+4\lambda
z^{2}}}\left[\sigma^{(L)}(z,\lambda)+2\lambda z^{2}\sigma^{(2)}(z,\lambda)\right]. \label{36.3}
\end{eqnarray}
Now we are able to compare our results with original HM ones. It is easy to see that the LO cross
sections (defined by Eqs.~(37) in \cite{HM} and Eqs.~(\ref{22.2}) in our paper) obey both above
identities.  Comparing with each other the quantities $\sigma^{(2)}_{1}$ and
$\hat{\sigma}_{2}^{(1)}$ (given by Eq.~(51) in \cite{HM} and Eq.~(\ref{25.2}) in this paper,
respectively), we find that identity (\ref{35.3}) is satisfied at NLO too. The situation with
longitudinal cross sections is more complicated. We have uncovered two misprints in the NLO
expression for $\sigma^{(L)}$ given by Eq.~(52) in \cite{HM}. First, the r.h.s. of this Eq. must be
multiplied by $z$. Second, the sign in front of the last term (proportional to $\delta (1-z)$) in
Eq.~(52) in Ref.~\cite{HM} must be changed \footnote{Note that this term originates from virtual
corrections and the virtual part of the longitudinal cross section given by Eq.~(39) in
Ref.~\cite{HM} also has wrong sign.}. Taking into account these typos, we find that relation
(\ref{36.3}) holds at NLO as well. So, our calculations of $\hat{\sigma}_{2}$ and
$\hat{\sigma}_{L}$ agree with the HM results.

Recently, the heavy quark initiated contributions to the $\varphi$-independent DIS structure
functions, $F_{2}$ and $F_{L}$, have been calculated by Kretzer and Schienbein (KS) \cite{KS}. The
final KS results are expressed in terms of the parton level structure functions $\hat{H}^{q}_{1}$
and $\hat{H}^{q}_{2}$. Using the definition of $\hat{H}^{q}_{1}$ and $\hat{H}^{q}_{2}$ given by
Eqs.~(7,8) in Ref.~\cite{KS}, we obtain that
\begin{equation}\label{37.3}
\hat{\sigma}_{T}(z,\lambda)\equiv
\frac{\alpha_{s}}{2\pi}\frac{\hat{\sigma}_{B}(z)}{\sqrt{1+4\lambda}}\frac{\hat{H}^{q}_{1}(\xi^{\prime},
\lambda)}{\sqrt{1+4\lambda z^{2}}},\qquad \qquad \qquad \qquad \hat{\sigma}_{2}(z,\lambda)\equiv
\frac{\alpha_{s}}{2\pi}\hat{\sigma}_{B}(z)\sqrt{\frac{1+4\lambda}{1+4\lambda
z^{2}}}\,\hat{H}^{q}_{2}(\xi^{\prime},\lambda),
\end{equation}
where $\hat{\sigma}_{T}=\hat{\sigma}_{2}-\hat{\sigma}_{L}$ and $\hat{\sigma}_{L}$ are defined by
Eq.~(\ref{27}) in our paper and
$\xi^{\prime}=z\left(1+\sqrt{1+4\lambda}\right)\left/\left(1+\sqrt{1+4\lambda
z^{2}}\right)\right.$. To test identities (\ref{37.3}), one needs only to rewrite the NLO
expressions for the functions $\hat{H}^{q}_{1}(\xi^{\prime},\lambda)$ and
$\hat{H}^{q}_{2}(\xi^{\prime},\lambda)$ (given in Appendix C in Ref.~\cite{KS}) in terms of
variables $z$ and $\lambda$. Our analysis shows that relations (\ref{37.3}) hold at both LO and
NLO. Hence we coincide with the KS predictions for the $\gamma^{*}Q$ cross sections.

However, we disagree with the conclusion of Refs.~\cite{KS,KS-thesis} that there are errors in the
NLO expression for $\sigma^{(2)}$ given in Ref.~\cite{HM}. As explained above, a correct
interpretation of the quantities $\sigma^{(2)}$ and $\sigma^{(L)}$ used in \cite{HM} leads to a
complete agreement between the HM, KS and our results for $\varphi$-independent cross sections.

As to the $\varphi$-dependent DIS, pQCD predictions for the $\gamma^{*}Q$ cross sections
$\hat{\sigma}_{A}(z,\lambda)$ and $\hat{\sigma}_{I}(z,\lambda)$ in the case of arbitrary values of
$m^{2}$ and $Q^{2}$ are not, to our knowledge, available in the literature. For this reason, we
have performed several cross checks of our results against well known calculations in two limits:
$m^{2}\rightarrow 0$ and $Q^{2}\rightarrow 0$. In particular, in the chiral limit, we reproduce the
original results of Georgi and Politzer \cite{GP} and M\'{e}ndez \cite{Mendez} for
$\hat{\sigma}_{I}(z,\lambda\rightarrow 0)$ and $\hat{\sigma}_{A}(z,\lambda\rightarrow 0)$. In the
case of $Q^{2}\rightarrow 0$, our predictions for $\hat{\sigma}_{2}(s,Q^{2}\rightarrow 0)$ and
$\hat{\sigma}_{A}(s,Q^{2}\rightarrow 0)$ given by Eqs.~(\ref{25.2},\ref{27.2}) reduce to the QED
textbook results for the Compton scattering of polarized photons \cite{Fano}.

\section{\label{IV}Some Properties of the Azimuth-Dependent Cross Sections}
To perform a numerical investigation of the inclusive partonic cross sections, $\hat{\sigma}_{i}$
($i=2,L,A,I$),{\large \ } it is convenient to introduce the dimensionless coefficient functions
$c_{i}^{(n,l)}$,
\begin{equation}\label{32.4}
\hat{\sigma}_{i}(\eta ,\lambda ,\mu ^{2})=\frac{e_{Q}^{2}\alpha _{em}\alpha _{s}(\mu
^{2})}{m^{2}}\sum_{n=0}^{\infty }\left( 4\pi \alpha _{s}(\mu ^{2})\right)
^{n}\sum_{l=0}^{n}c_{i}^{(n,l)}(\eta ,\lambda )\ln ^{l}\left( \frac{\mu ^{2}}{m^{2}}\right),
\end{equation}
where $\mu$ is a factorization scale (we use $\mu=\mu_{F}=\mu_{R}$) and the variable $\eta$
measures the distance to the partonic threshold:
\begin{equation}\label{33.4}
\eta =\frac{s}{m^{2}}-1=\frac{1-z}{\lambda z},\qquad \qquad \qquad \qquad \qquad s =(q+k)^{2}.
\end{equation}

\begin{figure}
\begin{center}
\begin{tabular}{cc}
\mbox{\epsfig{file=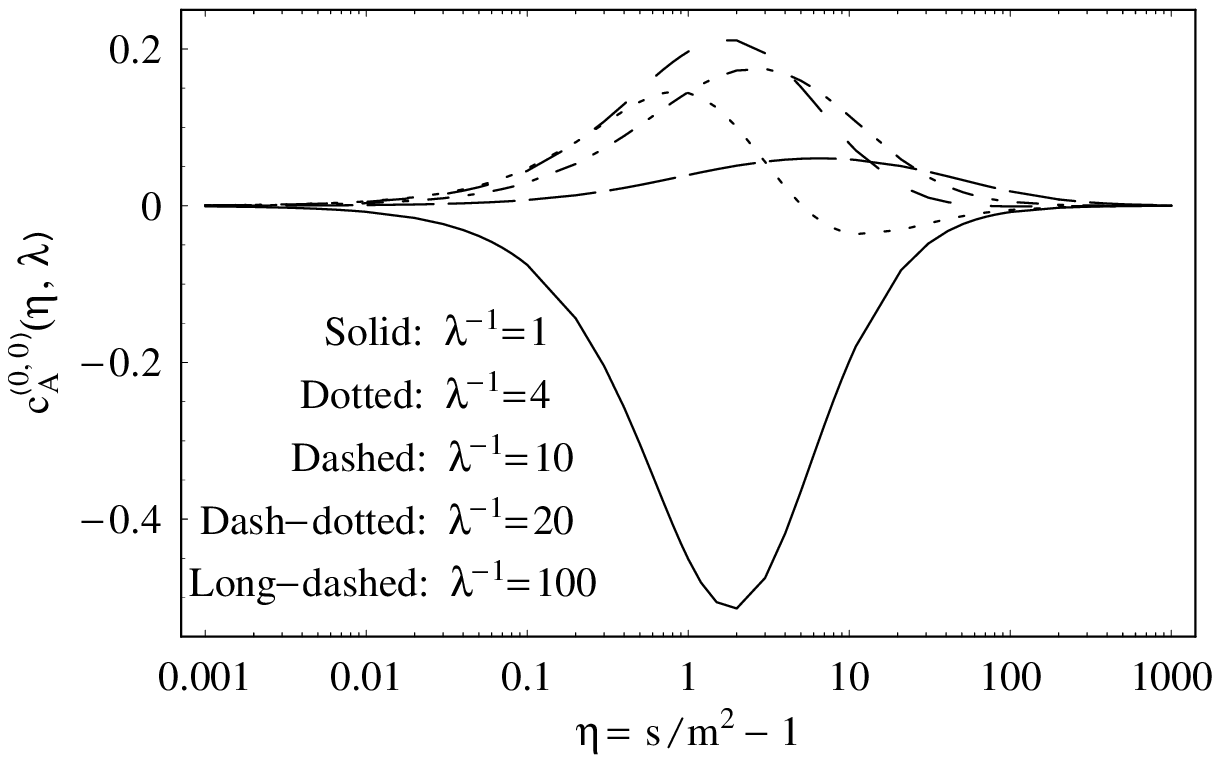,width=250pt}}
& \mbox{\epsfig{file=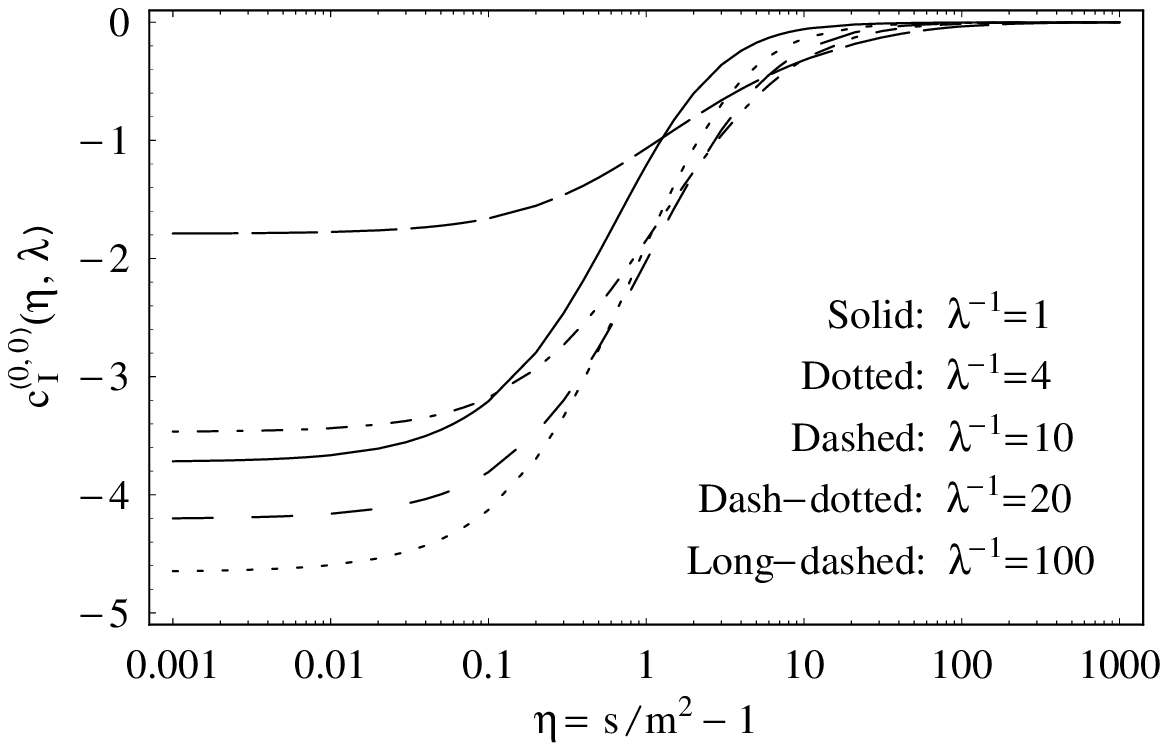,width=253pt}}\\
\end{tabular}
\caption{\label{Fg.3}\small  $c_{A}^{(0,0)}(\eta,\lambda )$ and $c_{I}^{(0,0)}(\eta,\lambda )$
coefficient functions at several values of $\lambda$.}
\end{center}
\end{figure}

Our analysis of the quantity $c_{A}^{(0,0)}(\eta ,\lambda)$ is given in Fig.~\ref{Fg.3}. One can
see that $c_{A}^{(0,0)}$ is negative at low $Q^{2}$ ($\lambda^{-1}\lesssim 1$) and positive at high
$Q^{2}$ ($\lambda^{-1}> 20$). For the intermediate values of $Q^{2}$, $c_{A}^{(0,0)}(\eta
,\lambda)$ is an alternating function of $\eta$.

Let us discuss the coefficient function $c_{A}^{(0,0)}(s,Q^{2})$ for the case of on-mass-shell
photon, $Q^{2}\rightarrow 0$. In this limit,
\begin{equation}\label{35.4}
c_{A}^{(0,0)}(s,Q^{2}\rightarrow 0)=8\pi C_{F}\frac{m^{4}}{(s-m^{2})^{2}}\left[
2+\frac{s+m^{2}}{s-m^{2}}\ln\frac{m^{2}}{s}\right]+ {\cal{O}}(Q^{2}).
\end{equation}
Considering now the threshold behavior of Eq.~(\ref{35.4}), we find:
$\displaystyle{\lim_{s\rightarrow m^{2}}}c_{A}^{(0,0)}(s,Q^{2}\rightarrow 0)=-4\pi C_{F}/3$. Taking
also into account that $\displaystyle{\lim_{s\rightarrow m^{2}}}c_{A}^{(0,0)}(s,Q^{2}\neq 0)=0$, we
see that the mass-shell, $Q^{2}\rightarrow 0$, and threshold, $s\rightarrow m^{2}$, limits do not
commutate with each other for the quantity $c_{A}^{(0,0)}(s,Q^{2})$. This property of the cross
section $c_{A}^{(0,0)}(s,Q^{2})$ illustrates the well-known fact that there is no, generally
speaking, a smooth transition between the lepto- and photoproduction.

Our results for the coefficient function $c_{I}^{(0,0)}(\eta,\lambda )$ at several values of
$\lambda$ are presented in Fig.~\ref{Fg.3}. It is seen that $c_{I}^{(0,0)}$ is negative at all
values of $\eta$ and $\lambda$. Note also the threshold behavior of the coefficient function:
\begin{equation}\label{34.4}
c_{I}^{(0,0)}(\eta\rightarrow 0,\lambda )=-\sqrt{2}\,\pi^{2}C_{F}\frac{\sqrt{\lambda}}{1+4\lambda}+
{\cal{O}}(\eta).
\end{equation}
This quantity takes its minimum value at $\lambda_{m}=1/4$: $c_{I}^{(0,0)}(\eta = 0,\lambda_{m})
=-\pi^{2}C_{F}/\left(2\sqrt{2}\right)$.

In the chiral limit, $m^{2}\rightarrow 0$, the $\varphi$-dependent cross section are as follows:
\begin{equation}\label{36.4}
c_{A}^{(0,0)}(z,\lambda\rightarrow 0)=2\pi C_{F}\lambda z^{2}+ {\cal{O}}(\lambda^{2}),\qquad \qquad
\qquad c_{I}^{(0,0)}(z,\lambda\rightarrow 0)=-\frac{\sqrt{2}}{4}\pi^{2}C_{F}\lambda
z(1+2z)\sqrt{\frac{z}{1-z}}+ {\cal{O}}(\lambda^{2}).
\end{equation}

Let us analyze the numerical significance of the $\cos\varphi$- and $\cos2\varphi$-distributions
for the QS component. It is difficult to compare directly the $\hat{\sigma}^{(1)}_{A}(z,\lambda)$
and $\hat{\sigma}^{(1)}_{I}(z,\lambda)$ cross section given by the usual functions (\ref{27.2}) and
(\ref{28.2}) with the $\varphi$-independent contributions $\hat{\sigma}^{(0)}_{2}(z,\lambda)$ and
$\hat{\sigma}^{(1)}_{2}(z,\lambda)$ described by the generalized functions (\ref{22.2}) and
(\ref{25.2}). For this reason, we consider the Mellin moments of the corresponding quantities
defined as
\begin{equation}\label{37.4}
\hat{\sigma}_{i}(N,\lambda)=\int\limits^{1}_{0}\hat{\sigma}_{i}(z,\lambda)z^{N-1}\text{d}z, \qquad
\qquad \qquad (i=2,L,A,I).
\end{equation}
The Mellin transform of the Born level cross sections is trivial:
$\hat{\sigma}^{(0)}_{2}(N,\lambda)=\hat{\sigma}_{B}(1)\sqrt{1+4\lambda}$. The Mellin moments of the
NLO results have been calculated numerically. We use for $\alpha_{s}(\mu_{F})$ the one-loop
approximation with $\Lambda_{4}=326$ MeV, $\mu_{F}=\sqrt{m^{2}+Q^{2}}$ and $m=1.3$ GeV.
\begin{figure}
\begin{center}
\begin{tabular}{cc}
\mbox{\epsfig{file=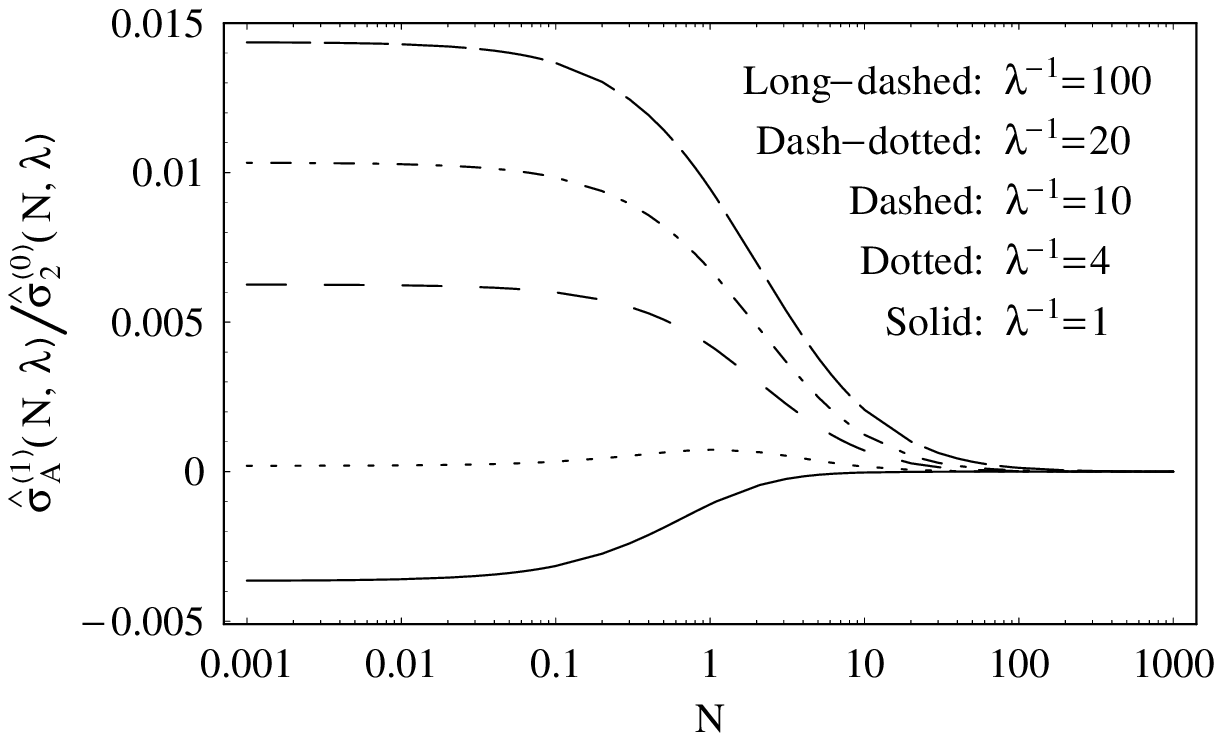,width=250pt}}
& \mbox{\epsfig{file=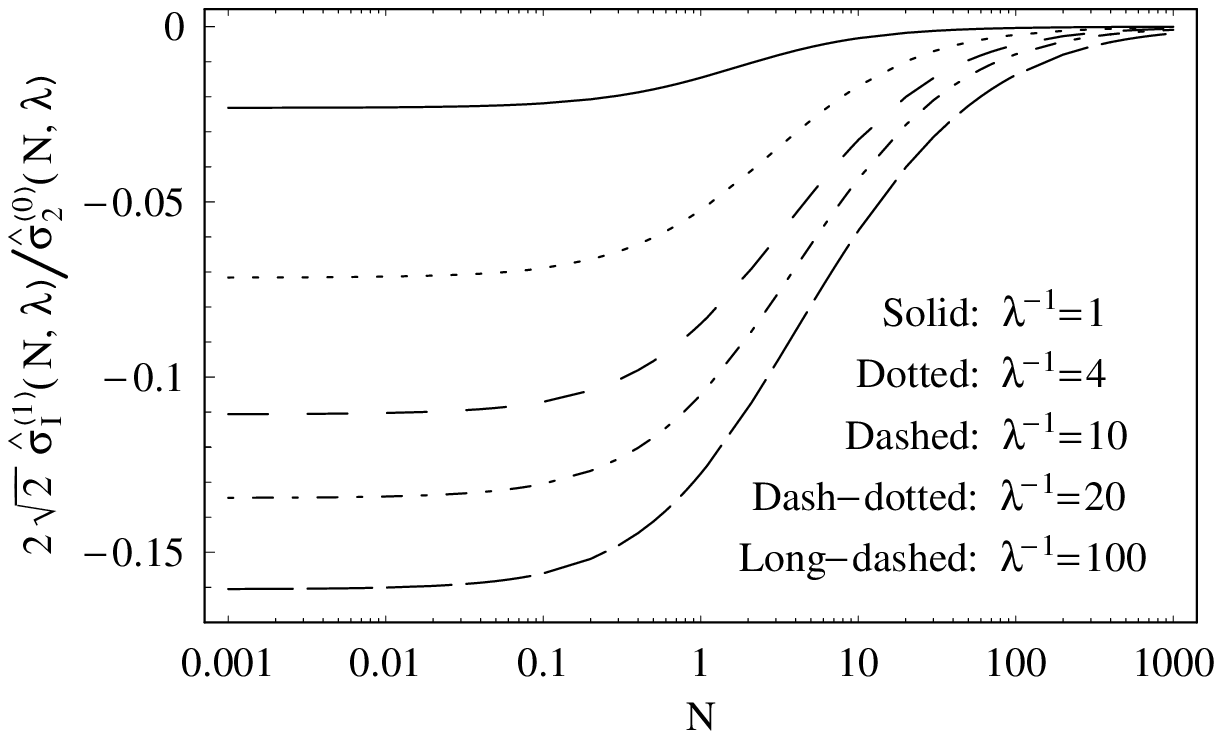,width=247pt}}\\
\end{tabular}
\caption{\label{Fg.4}\small The quantities
$\hat{\sigma}^{(1)}_{A}(N,\lambda)/\hat{\sigma}^{(0)}_{2}(N,\lambda)$ (\emph{left panel}) and
$2\sqrt{2}\,\hat{\sigma}^{(1)}_{I}(N,\lambda)/\hat{\sigma}^{(0)}_{2}(N,\lambda)$ (\emph{right
panel}) at several values of $\lambda$.}
\end{center}
\end{figure}

The left panel of Fig.~\ref{Fg.4} presents the ratio
$\hat{\sigma}^{(1)}_{A}(N,\lambda)/\hat{\sigma}^{(0)}_{2}(N,\lambda)$ as a function of $N$ for
several values of variable $\lambda$: $\lambda^{-1}=1,4,10,20$ and 100. One can see that this ratio
is negligibly small (of the order of 1$\%$). Moreover, our analysis shows that the ratio
$\hat{\sigma}^{(1)}_{A}(N,\lambda)/\hat{\sigma}^{(0)}_{2}(N,\lambda)$ is less than $1.5\%$ for all
values of $\lambda$ and $N>0$. This implies that the photon-quark scattering contribution is
practically $\cos2\varphi$-independent.

In the right panel of Fig.~\ref{Fg.4}, the $N$-dependence of the ratio
$2\sqrt{2}\,\hat{\sigma}^{(1)}_{I}(N,\lambda)/\hat{\sigma}^{(0)}_{2}(N,\lambda)$ is given for the
same values of $\lambda$. One can see that this ratio is of the order of 10-15$\%$ at small $N$ and
sufficiently high $Q^{2}$. This fact indicates that the $\cos\varphi$-distribution caused by the QS
component may be sizable.

\section{\label{V}Conclusion}
We conclude by summarizing our main observations. In the present paper, we have studied the
azimuth-dependent photon-(heavy) quark DIS at NLO. It turns out that the $\cos2\varphi$ dependence
of the QS mechanism is negligible while the $\cos\varphi$ one may be sizable. The situation is
diametrically opposite to the one that takes place for the basic GF contribution. It is well known
that the GF predictions for the azimuthal $\cos2\varphi$ asymmetry in heavy quark photo-
\cite{Duke-Owens,we1} and leptoproduction \cite{LW1,Watson,we4} are large (about 20$\%$). As to the
$\cos\varphi$ dependence of the GF contribution, it vanishes at LO due to the charge symmetry
$Q\leftrightarrow\overline{Q}$ \cite{LW2}.

Since the GF and QS mechanisms have strongly different azimuthal distributions, one could expect
that measurements of the $\varphi$-dependent DIS will directly probe the charm content of the
proton. In detail, hadron level predictions for the azimuthal asymmetries as well as the
possibility to discriminate experimentally between the GF and QS contributions will be investigated
in Ref.~\cite{we5}.
\begin{acknowledgments}
We thank S.J. Brodsky for stimulating discussions and useful suggestions. We also would like to
acknowledge interesting correspondence with I. Schienbein. This work was supported in part by the
ANSEF grants 04-PS-hepth-813-98, PS-condmatth-521 and NFSAT grant GRSP-16/06.
\end{acknowledgments}

\end{document}